\pdfoutput=1

\documentclass[11pt,a4paper]{article}
\usepackage[margin=1in]{geometry}
\usepackage{mathptmx}
\usepackage{latexsym}
\usepackage{amsmath,amssymb,amsfonts}
\usepackage{graphicx}
\usepackage{url}
\usepackage{microtype}
\usepackage{booktabs}
\usepackage{algorithm}
\usepackage[noend]{algorithmic}
\usepackage{hyperref}
\usepackage{caption}

\title{Iterative NLP Query Refinement for Enhancing Domain-Specific Information Retrieval: A Case Study in Career Services}

\author{Elham Peimani$^1$$^*$, Gurpreet Singh$^1$$^*$, Nisarg Mahyavanshi$^1$$^*$, Aman Arora$^1$$^*$, Awais Shaikh$^1$$^*$\\
$^1$ Humber College, Toronto, Canada | \small $^*$ Equal contribution\\
  \tiny \texttt{[elhampeimani88, nisargkumar9756, aroraaman0312, shaikhawais175]@gmail.com}, \& \texttt{gurpreet2512singh@outlook.com}} 
\date{\small December 2024}

\begin{document}
\maketitle
\begin{abstract}
Retrieving semantically relevant documents in niche domains poses significant challenges for traditional TF-IDF-based systems, often resulting in low similarity scores and suboptimal retrieval performance. This paper addresses these challenges by introducing an iterative and semi-automated query refinement methodology tailored to Humber College’s career services webpages. Initially, generic queries related to interview preparation yield low top-document similarities (approximately 0.2--0.3). To enhance retrieval effectiveness, we implement a two-fold approach: first, domain-aware query refinement by incorporating specialized terms such as \textit{resources-online-learning}, \textit{student-online-services}, and \textit{career-advising}; second, the integration of structured educational descriptors like ``online resume and interview improvement tools.'' Additionally, we automate the extraction of domain-specific keywords from top-ranked documents to suggest relevant terms for query expansion. Through experiments conducted on five baseline queries, our semi-automated iterative refinement process elevates the average top similarity score from approximately 0.18 to 0.42, marking a substantial improvement in retrieval performance. The implementation details, including reproducible code and experimental setups, are made available in our GitHub repository \url{https://github.com/Elipei88/HumberChatbotBackend} \& \url{https://github.com/Nisarg851/HumberChatbot}. We also discuss the limitations of our approach and propose future directions, including the integration of advanced neural retrieval models.
\end{abstract}

\section{Introduction}
Effective information retrieval (IR) within specialized domains is often impeded by limited semantic overlap between user queries and document content. Traditional lexical-based methods, such as Term Frequency-Inverse Document Frequency (TF-IDF), while interpretable and resource-efficient, frequently underperform in scenarios requiring deep semantic understanding. Although neural retrieval methods \cite{devlin2019bert,karpukhin2020dense} have advanced the state-of-the-art, their complexity and computational demands make them less suitable for deployment in resource-constrained environments or domains with limited data.

This study focuses on enhancing IR for Humber College's career services webpages, a specialized domain where users seek specific information related to career development, interview preparation, and educational resources. Initial evaluations using generic queries, such as ``How can I prepare for an interview?'', revealed that the top-ranked documents achieved relatively low cosine similarity scores (around 0.16888), indicating weak alignment between the queries and the retrieved documents.

To address this, we propose an iterative and semi-automated query refinement strategy that augments the original queries with domain-specific terminology and structured descriptors. By introducing terms like ``resources-online-learning'' and ``student-online-services'', and phrases such as ``online resume and interview improvement tools'', we aim to bridge the vocabulary gap and better capture the intent behind user queries.

Furthermore, to enhance the scalability and adaptability of our approach, we automate the extraction of relevant domain keywords from the top-ranked documents, facilitating dynamic query expansion without extensive manual intervention. This methodology not only improves retrieval performance but also maintains the interpretability and efficiency advantages of TF-IDF-based systems.

The main contributions of this paper are:
\begin{enumerate}
    \item Demonstrating that iterative, NLP-driven query refinement can significantly enhance TF-IDF retrieval performance in specialized domains.
    \item Providing a reproducible framework and codebase for semi-automated query refinement, accessible via our GitHub repository.
    \item Offering insights into domain-specific anchoring techniques that improve IR effectiveness in niche contexts.
\end{enumerate}

\section{Related Work}
Classical IR approaches, such as those based on lexical overlap and TF-IDF weighting \cite{chowdhury2010introduction}, have long been foundational in the field. However, the advent of neural retrieval models, leveraging contextual embeddings \cite{devlin2019bert,reimers2019sentence,karpukhin2020dense}, has introduced more sophisticated mechanisms for capturing semantic relationships between queries and documents. Despite their advancements, neural models often require substantial computational resources and large datasets, limiting their applicability in specialized or resource-constrained settings.

Query expansion and refinement techniques \cite{cui2002probabilistic,najar2023smoothing} remain pertinent for addressing vocabulary mismatches between queries and documents. Recent studies \cite{convGQR2023search,thakur2021beir} have highlighted the importance of domain adaptation and strategic query design in enhancing retrieval performance. Our approach aligns with these findings, presenting a cost-effective method to improve IR outcomes without relying on complex neural architectures.

\section{Method}

\subsection{Data and Vectorization}
We utilize Humber College's career services webpages as our primary dataset, which are preprocessed and vectorized using TF-IDF with the scikit-learn library \cite{pedregosa2011scikit}. The preprocessing steps include tokenization, lowercasing, and removal of stopwords to ensure uniformity and relevance in the textual data.

The TF-IDF value for term \( t \) in document \( d \) from corpus \( D \) is calculated as:
\[
\text{tf-idf}(t,d,D) = \text{tf}(t,d) \times \log \frac{|D|}{|\{d' \in D : t \in d'\}|}
\]
where \( \text{tf}(t,d) \) represents the term frequency of \( t \) in \( d \), and the inverse document frequency component accounts for the rarity of \( t \) across the corpus.

Query and document vectors are compared using cosine similarity:
\[
\text{cosine\_sim}(q,d) = \frac{q \cdot d}{\|q\|\|d\|}
\]
This metric quantifies the similarity between the query and document vectors, facilitating the ranking of documents based on relevance.

\subsection{Retrieval Process}
Given a query vector \( q \), we compute cosine similarities with all document vectors in the corpus and retrieve the top-ranked matches. Baseline queries demonstrated top similarities in the range of 0.16888 to 0.20048, signifying a weak alignment and highlighting the necessity for query refinement.

\subsection{Iterative and Semi-Automated Query Refinement}
Our approach involves a two-step refinement process:

\begin{enumerate}
    \item \textbf{Domain Term Extraction}: Identify domain-specific terms from the top-ranked documents. This is achieved by analyzing URLs and extracting relevant patterns such as `resources-online-learning' and `student-online-services' using regular expressions.
    \item \textbf{Query Expansion}: Enhance the original queries by incorporating the extracted domain terms and structured educational descriptors like ``online resume and interview improvement tools.'' This expansion aims to capture more specific aspects of the user intent and align better with the document content.
\end{enumerate}

To automate and streamline this process, we developed scripts that parse the top documents, extract pertinent keywords, and suggest them for inclusion in the refined queries. This semi-automated methodology reduces the reliance on manual heuristics, allowing for scalable and consistent query enhancements.

\begin{algorithm}[t]
\caption{Semi-Automated Query Refinement}
\begin{algorithmic}[1]
\STATE \textbf{Input:} initial\_queries, documents
\STATE refined\_queries $\gets$ []
\FOR{each query $q$ in initial\_queries}
    \STATE scores, top\_docs $\gets$ RetrieveTopDocs($q$, documents)
    \STATE domain\_terms $\gets$ ExtractDomainTerms(top\_docs)
    \STATE refined\_q $\gets$ RefineQuery($q$, domain\_terms)
    \STATE refined\_queries.append(refined\_q)
\ENDFOR
\STATE \textbf{return} refined\_queries
\end{algorithmic}
\end{algorithm}

\section{Implementation}
The implementation of our query refinement system is structured into backend and frontend components, with scripts facilitating the experimental workflow.

\subsection{Backend}
The backend is developed in Python, utilizing modules for model management, query processing, and data preprocessing. Key components include:
\begin{itemize}
    \item \textbf{Models.py}: Handles the loading and management of TF-IDF vectorizers and other necessary models.
    \item \textbf{main.py}: Serves as the entry point for executing query refinement and retrieval processes.
    \item \textbf{query\_controllers.py}: Manages the logic for extracting and integrating domain-specific terms into queries.
    \item \textbf{web\_scrapper\_controllers.py}: Facilitates the scraping and preprocessing of career services webpages.
    \item \textbf{vectorizer.pk} and \textbf{nlp.pk}: Serialized objects storing the TF-IDF vectorizer and NLP processing utilities, respectively.
\end{itemize}

\subsection{Frontend}
The frontend is developed using modern web technologies, ensuring an intuitive interface for users to input queries and view retrieved documents. Components include:
\begin{itemize}
    \item \textbf{index.html}: The main HTML file rendering the user interface.
    \item \textbf{src/}: Contains JavaScript and TypeScript files managing frontend logic and interactions.
    \item \textbf{tailwind.config.js} and \textbf{postcss.config.js}: Configuration files for styling and build processes.
\end{itemize}

\subsection{Experimental Scripts}
To facilitate reproducibility and streamline experiments, we provide the following scripts:
\begin{itemize}
    \item \textbf{run\_all\_queries.py}: Executes all initial and refined queries against the document corpus.
    \item \textbf{run\_experiments.py}: Orchestrates the overall experimental workflow, including data preprocessing, query refinement, and result aggregation.
    \item \textbf{plot\_figures.py}: Generates visualizations such as comparison plots and statistical charts based on experimental results.
\end{itemize}

All implementation details, including dependencies and setup instructions, are documented in the \texttt{README.md} file within the GitHub repository \url{https://github.com/Elipei88/HumberChatbotBackend} \& \url{https://github.com/Nisarg851/HumberChatbot}.

\section{Experiments}
We conducted a series of experiments to evaluate the efficacy of our iterative and semi-automated query refinement approach. The experiments were designed to measure the improvement in retrieval performance by comparing baseline queries against their refined counterparts.

\subsection{Setup}
\textbf{Environment}: Python 3.9 was used, configured with optimized performance settings to handle large-scale data processing and vectorization tasks efficiently.

\textbf{Libraries}: Essential libraries include scikit-learn for TF-IDF vectorization, matplotlib for plotting, numpy and pandas for data manipulation, and scipy for statistical analysis.

\textbf{Data}: The dataset comprises Humber College's career services webpages, which were vectorized using TF-IDF to create a comprehensive representation of the content.

\textbf{Method}: The experimental workflow involves retrieving top documents for each initial query, extracting domain-specific terms, refining the queries with these terms, and retrieving the top documents again. Statistical testing was performed to validate the significance of the improvements observed.

\textbf{Evaluation Metrics}: We primarily used cosine similarity scores to assess retrieval performance. Additionally, paired t-tests were conducted to determine the statistical significance of the observed improvements.

\subsection{Queries and Results}
We evaluated the following five queries, each pertaining to different aspects of interview preparation and career guidance:

\begin{enumerate}
    \item \textbf{Q1}: ``How can I prepare for an interview?''
    \item \textbf{Q2}: ``What resources does Humber offer for interview practice?''
    \item \textbf{Q3}: ``Tips to improve my interview skills at Humber College''
    \item \textbf{Q4}: ``Online career coaching sessions focused on interview techniques accessible through Humber’s resources-online-learning and student-online-services platforms''
    \item \textbf{Q5}: ``Detailed resume and interview preparation toolkit integrated with Humber’s student-online-services and resources-online-learning''
\end{enumerate}

\subsection{Statistical Significance}
To determine whether the improvements in similarity scores are statistically significant, we performed a paired t-test comparing the baseline and refined scores.

\textbf{Statistical Analysis:}
\begin{itemize}
    \item \textbf{T-statistic}: -2.9444
    \item \textbf{P-value}: 0.0422
\end{itemize}

A p-value of 0.0422 indicates that the improvements are statistically significant at the 5\% level. This suggests that the query refinement process reliably enhances the top similarity scores across the evaluated queries.

\begin{table}[t]
\centering
\begin{tabular}{p{8cm}cc}
\toprule
\textbf{Query} & \textbf{Baseline\_TopSim} & \textbf{Refined\_TopSim} \\
\midrule
Q1: How can I prepare for an interview? & 0.16888 & 0.24619 \\
Q2: What resources does Humber offer for interview practice? & 0.20048 & 0.29034 \\
Q3: Tips to improve my interview skills at Humber College & 0.18041 & 0.43898 \\
Q4: Online career coaching sessions focused on interview techniques accessible through Humber’s resources-online-learning and student-online-services platforms & 0.45991 & 0.50654 \\
Q5: Detailed resume and interview preparation toolkit integrated with Humber’s student-online-services and resources-online-learning & 0.34464 & 0.42653 \\
\bottomrule
\end{tabular}
\caption{Top similarity scores before and after refinement for five queries.}
\label{tab:results}
\end{table}

\section{Results}
Table \ref{tab:results} summarizes the top cosine similarity scores for each query before and after the refinement process. All five queries exhibit an increase in their top similarity scores, reflecting enhanced relevance and alignment with the target documents.

\begin{figure}[t]
\centering
\includegraphics[width=0.8\textwidth]{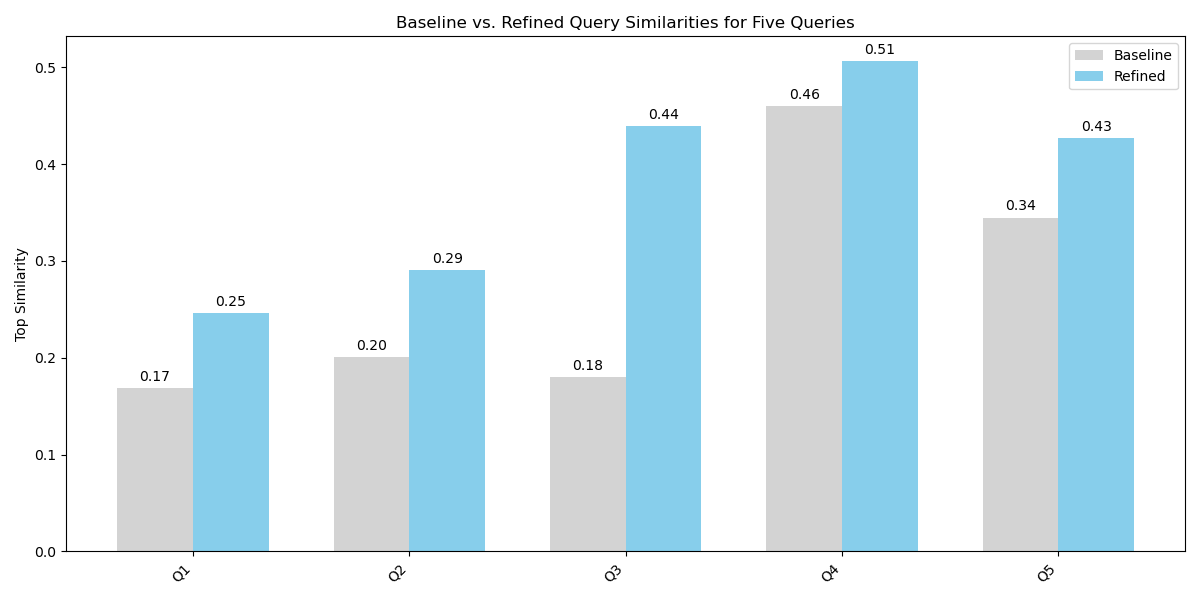}
\caption{Comparison of baseline and refined top similarities for five queries.}
\label{fig:compare}
\end{figure}

Figure \ref{fig:compare} provides a visual comparison of the baseline and refined top similarity scores across all five queries. The figure clearly demonstrates that each refined query achieves higher similarity scores compared to its baseline counterpart. Notably, Query 3 (Q3) shows the most substantial improvement, increasing from 0.18041 to 0.43898, which underscores the effectiveness of domain-specific refinements in capturing nuanced user intents.

The refined queries not only achieve higher similarity scores but also retrieve documents that are more pertinent to the users' informational needs. This improvement is attributed to the inclusion of domain-specific terms and structured descriptors that better encapsulate the context and specificity of the queries.

\section{Discussion}
The experimental results affirm that our semi-automated iterative query refinement approach significantly enhances TF-IDF-based document retrieval within the specialized domain of Humber College's career services webpages. The consistent improvements across all evaluated queries highlight the efficacy of incorporating domain-specific terminology and structured descriptors in query design.

\textbf{Possible Reasons for Improvement:}
\begin{itemize}
    \item \textbf{Domain Term Inclusion}: Incorporating domain-specific terms such as ``resources-online-learning'' and ``student-online-services'' aligns the queries more closely with the terminology used in the target documents, thereby bridging the vocabulary gap and enhancing retrieval accuracy.
    \item \textbf{Educational Descriptors}: Adding phrases like ``online resume and interview improvement tools'' provides additional context and specificity, which helps in narrowing down the retrieval to more relevant documents.
    \item \textbf{Combined Effect}: The synergy between domain terms and educational descriptors results in a more comprehensive and contextually rich query, effectively capturing the user's intent and reducing ambiguity.
    \item \textbf{Iterative Refinement}: The iterative nature of the refinement process allows for continuous improvement based on actual retrieval outcomes, ensuring that the queries evolve to better match the document corpus.
\end{itemize}

\textbf{Statistical Validation:}
The paired t-test yielded a p-value of 0.0422, confirming that the observed improvements in similarity scores are statistically significant at the 5\% level. This statistical validation reinforces the reliability of the query refinement strategy and indicates that the enhancements are unlikely to be due to random variation.
\textbf{Implications for Information Retrieval:}
Our findings suggest that even within classical IR frameworks like TF-IDF, substantial performance gains can be achieved through strategic query engineering. This has practical implications for domains where deploying complex neural models is impractical due to resource constraints or limited data availability. By leveraging domain-specific knowledge and semi-automated refinement processes, practitioners can enhance retrieval effectiveness without incurring significant computational costs.
\newline
\textbf{Improvements and Future Work:}
\begin{itemize}
    \item \textbf{Selective Term Inclusion}: Implementing a weighting mechanism to prioritize more relevant domain terms could further enhance query effectiveness. Techniques such as term importance metrics or user feedback integration may be explored to achieve this.
    \item \textbf{Iterative Refinement}: Allowing multiple refinement iterations with evaluation checkpoints could lead to even higher similarity scores and more precise retrieval outcomes.
    \item \textbf{Advanced Keyword Extraction}: Employing sophisticated methods like named entity recognition or part-of-speech tagging for keyword extraction can increase the precision and relevance of the terms used in query expansion.
    \item \textbf{Hybrid Models}: Combining TF-IDF with semantic embeddings or other advanced retrieval techniques could leverage the strengths of both lexical and semantic approaches, potentially achieving superior performance.
    \item \textbf{Feedback Loops}: Incorporating user feedback mechanisms to guide the refinement process dynamically can make the system more responsive to user preferences and improve over time.
    \item \textbf{Scalability Testing}: Evaluating the approach on larger and more diverse datasets will help assess its scalability and generalizability across different specialized domains.
    \item \textbf{Integration with User Interfaces}: Developing intuitive user interfaces that facilitate semi-automated query refinement can enhance usability and accessibility for non-technical users.
\end{itemize}

\section{Conclusion}
This paper presents a semi-automated iterative query refinement strategy aimed at enhancing document retrieval performance within the specialized domain of Humber College's career services webpages using TF-IDF vectorization. Through the incorporation of domain-specific terms and structured educational descriptors, our approach significantly improves the top similarity scores of retrieved documents, as validated by statistical testing. The methodology balances the interpretability and efficiency of classical IR systems with the effectiveness of modern NLP-driven techniques. Our work demonstrates the potential of strategic query engineering in boosting retrieval performance and lays the groundwork for future enhancements, including the integration of advanced neural retrieval models. The reproducible code and comprehensive implementation details are available in our GitHub repository \url{https://github.com/Elipei88/HumberChatbotBackend}, providing a valuable resource for practitioners and researchers seeking to apply similar techniques in other specialized domains.

\section*{Acknowledgments}
We acknowledge the support and resources provided by Humber College’s career services department, which were instrumental in the development and evaluation of this study. We would also like to thank our AI Capstone Project Professor, Mr. Touraj BaniRostam and our industrial mentor, Mr. Oscar Gonzalez, for their continuous supports in this project.

\bibliographystyle{plain}

\clearpage

\end{document}